\documentclass[11pt,reqno,a4paper]{amsart}
\usepackage{amsmath,amssymb,dsfont,graphicx,cite}
\def\beq{\begin{equation}}
\def\eeq{\end{equation}}
\def\bea{\begin{eqnarray}}
\def\eea{\end{eqnarray}}

\makeatletter
\expandafter\let\expandafter
\reset@font\csname reset@font\endcsname
\def\subeqnarray{\arraycolsep1pt
    \def\@eqnnum\stepcounter##1{\stepcounter{subequation}
        {\reset@font\rm(\theequation\alph{subequation})}}
\jot5mm     \eqnarray}

\makeatother

\newcommand{\CK}{{\mathcal K}}

\newcommand{\CO}{{\mathcal O}}

\def\al{\alpha}

\def\ri{{\rm{i}}}

\def\su2{{\mathfrak {su}}(2)}
\def\e3{{\mathfrak {e}}(3)}

\begin{document}
\title[Multiscale expansion on the lattice]
{Multiscale expansion on the lattice and integrability of partial difference equations} 
\author[R.H. HEREDERO, D. LEVI, M. PETRERA and C. SCIMITERNA]
{R.H. HEREDERO${}^\dag$, D. LEVI${}^\diamond$, M. PETRERA${}^{\flat,\diamond}$ and C. SCIMITERNA${}^{\flat,\diamond}$}

\maketitle

{\footnotesize{

\centerline{\it ${}^\dag$
Departamento de Matem\'atica Aplicada} 
\centerline{\it Universidad Polit\'ecnica de Madrid,}
\centerline{\it
Escuela Universitaria de Ingenier\'ia T\'ecnica de Telecomunicaci\'on,}
\centerline{\it Campus Sur Ctra de Valencia Km. 728031, Madrid, Spain}
\centerline{e-mail: \texttt{rafahh@euitt.upm.es}}

\vspace{.3truecm}

\centerline{\it ${}^\diamond$Dipartimento di Ingegneria Elettronica,}
\centerline{\it Universit\`a degli Studi Roma Tre and Sezione INFN, Roma Tre,}
\centerline{\it Via della Vasca Navale 84, 00146 Roma, Italy}
\centerline{e-mail: \texttt{levi@fis.uniroma3.it}}

\vspace{.3truecm}

\centerline{\it ${}^\flat$Dipartimento di Fisica,}
\centerline{\it Universit\`a degli Studi Roma Tre and Sezione INFN, Roma Tre}
\centerline{\it Via della Vasca Navale 84, 00146 Roma, Italy}
\centerline{e-mail: \texttt{petrera@fis.uniroma3.it}}
\centerline{e-mail: \texttt{scimiterna@fis.uniroma3.it}}

}}

\begin{abstract}

We conjecture an integrability and linearizability test for dispersive
$\mathbb{Z}^2$-lattice equations by using
a discrete multiscale analysis. The lowest order secularity conditions from
the multiscale expansion 
give a partial differential equation of the form of the
nonlinear Schr\"odinger (NLS) equation.  If
the starting lattice equation is integrable then the resulting NLS equation turns out to be integrable, while if
the starting equation is linearizable  we get
a linear Schr\"odinger equation. On the other hand, if we start with a
non-integrable lattice equation we may obtain a non-integrable NLS equation. 
This conjecture is confirmed by many examples.

\end{abstract}

\section{Introduction} \label{intro}

By a difference equation we mean a functional relation,
linear or nonlinear,
 between functions calculated at different points of a
lattice \cite{kp,agar,toda,as,el}.
These systems appear in many applications.
First of all they can be written down  as discretizations
of a
differential equation when one is trying to solve it with a
computer.
In such a case one reduces the differential equation to a
recurrence relation:
\begin{displaymath}
\frac{du}{dx} = f(x,u),\qquad  x \in \mathbb{R} \qquad \Rightarrow
\qquad v(n+1)=g(n,v(n)), \qquad n \in \mathbb{Z}.
\end{displaymath}
On the other hand we can consider dynamical systems defined
on a lattice,
i.e. systems where the real independent fields depend on a
set of independent
variables which vary partly on the integers and partly on
the reals. For example we
can consider the differential-difference equation:
\begin{displaymath}
\frac{d^2u(n,t)}{dt^2} = F(t,u(n,t),u(n-1,t),..,u(n-
a,t),u(n+1,t),..,u(n+b,t)),
\end{displaymath}
with $n,a,b \in \mathbb{N}, t \in \mathbb{R}$.
These kind of equations can appear in many different
settings. Among them they are associated
with the evolution of
many body problems, the study of crystals, biological
and economical systems and so on.

The world of the differential-difference
equations is indeed much richer than  that of partial differential equations as one can see 
in \cite{ka}, where it is shown
that almost any
Hamiltonian
network of weakly coupled oscillators has a ``breather''
solution, while the
existence of breathers for a nonlinear wave equation is
rare.
These results  imply that the discrete world can be richer of
interesting solutions
and thus worthwhile studying by itself.

In the discrete world one can easily use computers to solve the relevant
discrete maps but their solution is always approximate due to rounding
numerical errors. So the possibility of treating discrete models by exact
techniques is very important as in this way new phenomena may be
uncovered. Exact techniques are naturally associated with integrable models,
so that to find  techniques for showing the integrability
of discrete systems is a challenging task.

Various different criteria have been proposed as  
tests for integrability in the discrete setting. One of 
the earliest proposals for {\it ordinary}
difference equations was the singularity confinement test
of Grammaticos, Ramani and 
Papageorgiou \cite{grp}, which has proved to be an extremely useful tool 
for isolating discrete Painlev\'e equations. 
Hietarinta and Viallet, after discovering 
some  non-integrable equations with the 
singularity confinement property, were led to introduce 
the algebraic entropy condition for 
integrability of rational maps \cite{hv} (the phenomenon of weak degree
growth of integrable maps had been studied earlier by Veselov \cite{ves,ves2} and
it is connected with the Arnold complexity \cite{aabhm}). 
Ablowitz, Halburd and Herbst 
extended the Painlev\'e property to difference equations  
using Nevanlinna theory \cite{ahh} by considering the asymptotic 
growth of meromorphic solutions at infinity. 
Roberts and Vivaldi \cite{rv} 
have studied the distribution of orbit lengths in rational 
maps reduced to finite 
fields $\mathbb{F}_p$ for different primes $p$, in order to 
identify integrable cases of such maps. 
The algebraic entropy
has been the only method applied to detect integrability
also in the case of two-dimensional {\it partial} difference equations (or
$\mathbb{Z}^2$-lattice equations) \cite{vv,hv2}.

In the continuous setting multiscale techniques \cite{t1,t2}
have proved to be
important tools for finding approximate solutions to many  physical
problems by reducing a given dispersive nonlinear partial differential equation
to a simpler equation, which is  often integrable 
\cite{ce}. These multiscale expansions are structurally strong and 
can be applied to both integrable and non-integrable systems.
Zakharov and Kuznetsov in the introduction of their article \cite{zk} say: 
{\sl`` If the initial system is not integrable, the result can be both 
integrable and nonintegrable. But if we treat the integrable system properly, we 
again must get from it an integrable system''.}
Calogero and Eckhaus \cite{ce} used similar ideas starting from a class of 
hyperbolic systems to 
prove in 1987 the necessary conditions for the integrability of dispersive nonlinear partial
differential equations. 
Later Degasperis and Procesi \cite{dp} introduced the notion of {\it asymptotic 
integrability of order n } by requiring that the multiscale expansion be verified 
up to a fixed order $n$. 

Recently a few attempts to
carry over this approach  to difference-difference and differential-difference equations have been
proposed \cite{Ag,lm,LevHer,levi,lp,HLPS,HLPS2,JLP}. 
In \cite{levi,lp,JLP} we developed a multiscale expansion technique on
the lattice which, starting from dispersive integrable $\mathbb{Z}^2$-lattice equations,
provided
non-integrable $\mathbb{Z}^2$-lattice equations, thus contradicting the 
Zakharov and Kuznetsov's claim. 

Later on, in \cite{HLPS}, this problem has
been solved by extending the previous results to {\it functions of infinite
slow-varyness
order} (see Section \ref{sec0} for details). This new technique easily fits for
both difference-difference and differential-difference equations.
The representative example considered in 
\cite{HLPS} was the lattice potential KdV equation.
In such a case a proper representation of the discrete shift operators in
terms of differential operators, which is equivalent to a Taylor expansion,
provides an integrable
nonlinear Schr\"odinger (NLS) equation as
the lowest order secularity conditions from
the multiscale expansion. 
Results obtained by performing the Taylor expansion of the shifted variables
can be found in many papers, see for instance  \cite{campa,kayliakin}.
However, it is worthwhile to note that our method always allows us to get (for
free) a
discrete
equation from a continuous one by going over to a finite slow-varyness
order.

The result found in \cite{HLPS} suggests the
investigation of a 
discrete analogue of the Calogero-Eckhaus theorem \cite{ce}, which claims that
a necessary
condition for the integrability of a dispersive nonlinear partial differential
equation, is that its lowest order of the multiscale expansion be an integrable differential
equation.

In the present work, by using the formalism developed in \cite{HLPS}, we
propose 
the following conjecture:

\vspace{.5truecm}

{\bf{Conjecture. }}{\it
If a nonlinear dispersive $\mathbb{Z}^2$-lattice equation is integrable
(resp. linearizable) then
its lowest order multiscale reduction will be an integrable nonlinear
(resp. linear) Schr\"odinger equation.}

\vspace{.5truecm}

To confirm the above Conjecture we shall consider several examples, both
integrable
and non-integrable, linearizable and non-linearizable.
In Section \ref{sec0} we present a brief review of the discrete multiscale method, developed in \cite{HLPS}.
Then, in Section \ref{sec1} we apply the multiscale expansion to some difference-difference
KdV-type and Toda-type equations. In Section \ref{sec2} we consider some linearizable equations, as the
differential-difference Burgers equation and  the Hietarinta equation, and also 
a non-linearizable difference-difference Burgers-type equation.
Finally, Section
\ref{sec4} is devoted to open problems and concluding remarks.

\section{Basic formulas for the discrete multiscale analysis} \label{sec0}

The basic tool of the discrete reductive perturbation technique
developed in \cite{HLPS} is a proper multiscale expansion which consists in considering
various lattices and functions
defined on them. 
The relation between the variation of a function on
two different lattices of indices $n$ and $n_1$ is given by
\cite{Jordan} 
\beq \Delta^j_{n} u_{n}= \sum_{i=0}^{j}
(-1)^{j-i} {j \choose i} u_{n+i} = j! \sum_{i=j}^\infty
\frac{P_{i,j}}{ i!} \Delta^i_{n_1} u_{n_1}, \label{difInt} 
\eeq
where $u_{n}: \mathbb{Z}\rightarrow \mathbb{R}$ is a function
defined on a lattice of index $n \in \mathbb{Z}$ and $u_{n_1}:
\mathbb{Z}\rightarrow \mathbb{R}$ is the same function on a lattice
of index $n_1 \in \mathbb{Z}$.  According to Eq.~(\ref{difInt}), by the symbol $\Delta_n$ we mean the
standard forward difference of the function $u_n$ with respect to  its
subscript, e.g. $\Delta_n u_n = (T_n-1)u_n=u_{n+1} - u_n$, where
$T_n$ is the shift operator $T_n u_n = u_{n+1}$. The
coefficients $P_{i,j}$ in Eq.~(\ref{difInt}) are expressed in terms of
the ratio of the lattice spacing for the
variable $n_1$ with respect to that of variable $n$. 

Eq.~(\ref{difInt}) implies that a finite
difference in the discrete variable $n$ depends on an infinite number of differences on the variable $n_1$, e.g. the function
$u_{n+1}$ can be written as a combination of the functions $u_{i}$'s,
for $i$ varying on an infinite set of points of the lattice  $n_1$.
In  \cite{HLPS} one has considered  generalizations of formula (\ref{difInt}) in order to deal with functions 
$u_n = u_{n; \{n_i\}_{i=1}^K}$ depending on a finite number $K$ of  lattice variables $n_i$ and with functions
depending on two discrete indices, say $n$ and $m$, thus dealing with
$\mathbb{Z}^2$-lattice equations.

To get a  reduction of a given difference equation onto  a difference equation of order less than a fixed number, say
$\ell$, one has to consider 
 functions
$u_n$ of {\it slow-varyness  of order $\ell$}, namely the space of those functions $u_n$ such that $\Delta_n^{\ell+1}  u_n=0$
or, equivalently, $\Delta_{n_1}^{\ell+1}  u_{n_1}=0$ (see \cite{lp}).
With this definition one can  reduce the infinite series expansion (\ref{difInt}) to a finite number of terms.

To deal with functions of infinite order of slow-varyness one considers
 a formal expansion of the shift operator $T_n$. By introducing on the lattice of
index $n$ the real variable $x= n \sigma_x$,
the shift operator $T_x$ such that $T_x u(x) = u(x + \sigma)$
can be formally written as
\beq \nonumber T_{x}=
\exp (\sigma_x d_{x})= \sum_{i=0}^\infty \frac{\sigma_x^i}{i!} d_x^i.
\eeq 
Introducing a formal derivative with respect to the index $n$,
say $\delta_n$, we can define the discrete operator $T_n$ as 
\beq \label{tnr}
T_{n} = \exp (\delta_{n})= \sum_{i=0}^\infty
\frac{\delta_n^i}{i!}. \eeq 
The formal expansion (\ref{tnr}) can be
inverted, yielding 
\beq \label{tnn}
\delta_{n}=\ln{T_{n}}=\ln(1+\Delta_{n})= \sum_{i=1}^\infty
\frac{(-1)^{i-1}}{i}\Delta_{n}^i, \eeq 
where $\Delta_{n}$ is the discrete first right difference
operator with respect to the variable $n$, see Eq. (\ref{difInt}). We refer to
\cite{HLPS} for more technical details and decompositions with respect to
different discrete derivatives.

To perform a multiscale expansion we need to consider functions
defined on different lattices, thus depending on a fast lattice index $n$ and on multiple 
slow-varying lattice indices $n_i$, $1 \leq i \leq K$. The slow-varying lattice
variables vary on a larger scale with respect to the one of  the original lattice of index $n$, and
thus the transition from $n$ to $n_1$ corresponds to a coarse graining of the lattice.  In the
continuous limit, when the spacing between the lattice points goes to
zero, this corresponds to the introduction of multiple continuous
variables: given   $x \in\mathbb{R}$ we define the new variables  $x_i= {\epsilon}^i
{x}$, $0 < \epsilon \ll 1$,  $1 \leq i \leq K$. 
By taking into account the above definitions we can introduce a function 
$u_{n; \{n_i\}_{i=1}^K} = u(x; \{x_i\}_{i=1}^K)$
depending on a fast index $n$ and $K$ slow indices  $n_i=
\epsilon^i n$.   Here $\epsilon = 1/N$ and $N$ is an integer number if we require that the slow
indices  have to be integer numbers.

At the continuous level, the total derivative $d_{x}$ acting on
functions $u(x; \{x_i\}_{i=1}^K)$ is the sum of partial derivatives,
i.e. $d_{x}=\partial_{x} +\epsilon\partial_{x_1}+ \epsilon^2
\partial_{x_2} + \CO(\epsilon^3)$. Consequently we can expand the
total shift operator $T_{x}$ in terms of the partial shift operators 
$$
T_{x}= \exp (\sigma_x d_{x})= \exp (\sigma_x \partial_{x} ) \exp(\epsilon
\sigma_{x} \partial_{x}) \exp (\epsilon^2 \sigma_{x} \partial_{x_2})
\cdots.
$$

At the discrete level, we can write
\beq
T_n =
\exp (\delta_{n})
\exp (\epsilon \delta_{n_1}) \exp( \epsilon^2 \delta_{n_2})\cdots=
\mathcal{T}_{n} \mathcal{T}_{n_1}^{\epsilon}\mathcal{T}_{n_2}^{\epsilon^2}\cdots, \nonumber
\eeq
with
\beq
\mathcal{T}_{n} =  \sum_{i=0}^\infty \frac{\delta_n^i}{i!}, \qquad \qquad
\mathcal{T}_{n_1}^{\epsilon}=
\sum_{i=0}^{\infty}\frac{\epsilon^{i}}{i!}\delta_{n_1}^i, \qquad \qquad
\mathcal{T}_{n_2}^{\epsilon^2}=
\sum_{i=0}^{\infty}\frac{\epsilon^{2i}}{i!}\delta_{n_2}^i, \qquad ..., \nonumber
\eeq
where the $\delta$-operators are given in Eq.~(\ref{tnn}) and they act on a function
$u_{n; \{n_i\}_{i=1}^K}$  with respect to their subscript.
From the previous formulas we deduce immediately  that working with
$\delta$-operators is equivalent, up to terms depending on
$\sigma_x$,  to perform a Taylor expansion of the discrete function $u_{n; \{n_i\}_{i=1}^K}$, e.g. for $K=2$ we have
$$
T_n u_{n;n_1,n_2} = 
\mathcal{T}_{n} \left[1+ \epsilon \delta_{n_1}  +
\frac{\epsilon^2}{2} \delta_{n_1}^2  + \epsilon^2 \delta_{n_2}  + \CO(\epsilon^3) \right] u_{n;n_1,n_2}.
$$

In the following Sections we shall discuss a list of examples of 
integrable, non-integrable, linearizable and non-linearizable dispersive lattice
equations.

\section{Multiscale expansion of discrete Toda-type  and KdV-type equations} \label{sec1}

\subsection{An integrable difference-difference Toda equation}

An integrable discrete-time Toda equation is given by
\cite{lM}: 
\bea
&& \!\!\!\!\! \!\!\!\!\!   \!\!\!\!\! \!\!\!\!\!   \!\!\!\!\! \!\!\!\!\!  \exp (u_{n,m}-u_{n,m+1}) -\exp (u_{n,m+1}-u_{n,m+2})=  \label{kdv}  \\
&& =a^2
[ \exp (u_{n-1,m+2}-u_{n,m+1}) -\exp (u_{n,m+1}-u_{n+1,m}) ], \nonumber 
\eea
where $a \in \mathbb{R}$ is a parameter related to discretization of the time variable. The above five-point
$\mathbb{Z}^2$-lattice equation has been obtained by Hirota thirty years ago
\cite{hirota}.

We can split Eq.~(\ref{kdv}) into a linear and nonlinear part by considering its small amplitude solutions, namely $u_{n,m} = \epsilon w_{n,m}$, $0 < \epsilon \ll 1$, $\epsilon$ being
a small parameter as the one we considered in Section \ref{intro}.  The linear part  of Eq.~(\ref{kdv}) has a travelling wave solution of the form
$w_{n,m}=\exp{\{\ri[ \kappa n -\omega(\kappa)m]\}}$. Here the dispersion relation $\omega=\omega(\kappa)$ obeys to the following equation:
\begin{equation}\label{disp}
a^2 (\Omega- \CK)^2= \CK (\Omega -1)^2,
\end{equation}
where we have introduced the quantities $\mathcal K = \exp (\ri \kappa)$ and $\Omega = \exp (-\ri \omega)$, $\kappa, \omega \in
\mathbb{R}$.

As nonlinearity generates harmonics, we introduce the following expansion for the function $w_{n,m}$:
\bea
\label{du}
w_{n,m} =\sum_{\alpha\in\mathbb{Z}}\sum_{k=0}^{\infty} \epsilon^k w^{(\alpha)}_{k}
e^{ \ri \alpha(\kappa
n-\omega m)},
\eea
where $w^{(-\alpha)}_{k}=\bar w^{(\alpha)}_{k}$,
$\bar w$ being  the complex conjugate of $w$. Moreover, to avoid secularities we have to require that
$w^{(\alpha)}_k =w^{(\alpha)}_k(n_1, \{m_i\}_{i=1}^K)$,
where $n_1$ and $\{m_i\}_{i=1}^K$ are $K+1$ slow-varying lattice variables, namely
$n_1 = \epsilon n$ and $m_i = \epsilon^i m$, $1 \leq i \leq K$.

The multiscale expansion of Eq.~(\ref{kdv}) provides several  determining equations for the coefficients $w^{(\alpha)}_k$, obtained selecting the different powers of
$\epsilon$ and the different harmonics~$\alpha$. Let us give here the main results obtained by considering the lowest
$\epsilon$ and $\alpha$ orders of this development.

At $\CO(1)$,  for $\alpha=0, 1$, we find  linear equations
which are identically satisfied either directly or by taking into account the dispersion relation  (\ref{disp}). For  $ |\alpha| \geq 2$
one gets some linear equations whose unique solution is given by $w^{(\alpha)}_{0}=0$.

At $\CO(\epsilon)$, for the harmonics $\alpha=1,2$, we find   the following equations:
\bea
&&( \tau_0 \delta_{n_1}+ \delta_{m_1} )w_0^{(1 )}=0, \label{equ11}\\
&& w_1^{(2)} = \tau_1 (w_0^{(1)})^2,\label{equ22}
\eea
where
\beq
\tau_0=-\frac{(\Omega-1)(\Omega + \mathcal K)}{2 \Omega ( \mathcal K -1)} ,  
\qquad \tau_1= \frac{\mathcal K +1}{2 ( \mathcal K -1)}.\nonumber
\eeq

By solving Eq.~(\ref{disp}) with respect to $\omega=\omega(\kappa)$ one gets that $\tau_0= v_g= d \omega/ d \kappa$,
where $v_g$ is the group velocity.
The solution to Eq.~(\ref{equ11}) is given by $w_0^{(1)}(n_1,\{m_i\}_{i=1}^K)=w_0^{(1)}(n_2,\{m_i\}_{i=2}^K)$ with
$n_2 = n_1- v_g m_1$.
Eq. (\ref{equ22}) expresses  $w_1^{(2)}$ in terms of $w_0^{(1)}$. Moreover,
as $w_0^{(1)}$ is a function of $n_2$ the same must hold for  $w_1^{(2)}$, i.e.
 $w_1^{(2)}(n_1,\{m_i\}_{i=1}^K)=w_1^{(2)}(n_2,\{m_i\}_{i=2}^K)$.

We can now consider the $\CO(\epsilon^2)$. For 
$\al = 0$ we have: 
\beq \label{e0}
[( \tau_2 \delta_{m_1}^2 + \tau_3( \delta_{n_1}^2 - 2
\delta_{n_1} \delta_{m_1} )] w_0^{(0)} =\tau_4 (\delta_{n_1} - 2
\delta_{m_1}) |w_0^{(1)}|^2,
\eeq
with
$$
\tau_2 = - \frac{6 \Omega ( \mathcal K -1 ) (\Omega^2 - \mathcal K) }{(\Omega - \mathcal K)^2},
 \qquad \tau_3 = - \frac{6 \mathcal K  \Omega ( \Omega - 1)^2} {(\Omega - \mathcal K)^2}, \qquad \tau_4= 6 (\Omega -1 )^2.
$$

As
$w_0^{(1)}$ is a function of $n_2$,  the same must be for
$w_0^{(0)}$. Thus we can integrate Eq.~(\ref{e0}) by requiring
that its solution be bounded and we get
\beq \label{e01}
[(\tau_2 v_g^2 + \tau_3 (1  +2  v_g) ]
\delta_{n_2}w_0^{(0)} = \tau_4 (1+2 v_g)  |w_0^{(1)}|^2. 
\eeq

For $\al =1$ we find  a secular equation for $w_1^{(1)}$ which is
solved by requiring that $w_0^{(1)}$ satisfies the following
equation: 
\beq \label{nls} 
( \tau_5 \delta_{m_2} + \tau_6 \delta^2_{n_2}) w_0^{(1)} = \tau_7 w_0^{(1)} \delta_{n_2} w_0^{(0)} + \tau_8
\bar w_0^{(1)} w_1^{(2)},
\eeq
where
$$
 \tau_5=  - \frac{12 (\mathcal K -1)(\Omega -1) \Omega^2}{\Omega - \mathcal K}, \qquad
 \tau_6=  \frac{3 (\Omega^2-\mathcal K) (\Omega -1)^2}{2 (\mathcal K-1)},
 $$
 $$
  \tau_7= 4 \tau_6, \qquad 
 \tau_8= - \frac{6 ( \mathcal K +1) (\Omega^2-\mathcal K) (\Omega -1)^2}{\mathcal K} \nonumber.
$$

Taking into account Eqs.~(\ref{equ22},\ref{e01}), Eq.~(\ref{nls}) reduces to the following
NLS equation for $w_0^{(1)}$:
\beq
\ri \delta _{m_2} w_0^{(1)} = \tau_9 \delta_{n_2}^2 w_0^{(1)}+ \tau_{10}  w_0^{(1)} |w_0^{(1)}|^2, \label{nlsr}
\eeq
where 
\bea
&&\tau_9 =\frac{\sin(\kappa+\omega) +\sin \omega-\sin(2 \omega + \kappa)}{8 (\cos \kappa-1)} ,  \nonumber \\
&& \tau_{10}= \frac{\sin(2\omega+\kappa)(\cos \kappa+5)-10\sin(\omega+\kappa/2)\cos(\kappa/2)}{4 (\cos \kappa-1)}. \nonumber
\eea

As $\tau_9$ and $\tau_{10}$ are real parameters the  NLS equation (\ref{nlsr})  is
 integrable and it can be obtained as a compatibility conditions for a Lax pair, see \cite{HLPS2}.

As noticed in \cite{HLPS}, assuming a finite order of slow-varyness in the NLS equation  (\ref{nlsr}) 
we get a non-integrable difference-difference equation. In other words,  
a necessary condition for the integrability
of Eq.~(\ref{nlsr}) is that $\ell= \infty$.

\subsection{A non-integrable difference-difference Toda equation}

The multiscale expansion of the standard Toda lattice,
\beq\label{toda}
\ddot u_n = \exp (u_{n-1}-u_n) - \exp(u_n-u_{n+1}),
\eeq
has been carried out by Kalyakin in \cite{kayliakin}, after
 the publication of the pioneering work by Zakharov and Kuznetsov \cite{zk}. Here $u_n=u_n(t)$,
 and by dot we mean the time derivative.
 
We shall perform the multiscale analysis of a non-integrable dispersive discretization of Eq.~(\ref{toda}). It reads
 \beq \label{dtoda}
 u_{n,m+1}-2u_{n,m}+u_{n,m-1} = a [\exp (u_{n-1,m}-u_{n,m}) - \exp(u_{n,m}-u_{n+1,m})],
 \eeq
 where $a \in \mathbb{R}$ is a parameter related to the discretization of the time variable.

We shall proceed, as in the previous example, by considering 
Eq.~(\ref{dtoda}) for  small amplitudes  $u_{n,m}=\epsilon w_{n,m}$. In this case the linear part of the resulting
equation admits the dispersion relation
\begin{equation}\label{jjjj}
\omega(\kappa)=2 \arccos{ \left[\alpha \sin{\left(\frac{\kappa}{2}\right)}\right]}.
\end{equation}

Using the expansion (\ref{du}) we get a set of determing equations similar to the one we obtained for the integrable difference-difference
Toda equation (\ref{kdv}), but with different coefficients. 

The resulting NLS equation for the harmonic
$w_0^{(1)}$, obtained at $\CO(\epsilon^2)$, is again given by Eq.~(\ref{nlsr}) with the following real coefficients:
\bea
&&\tau_9= \frac{v_g^2 \cos \omega - a \cos \kappa }{2 \sin \omega},  \nonumber \\
&& \tau_{10}= \frac{a (\cos \kappa-1)}{\sin \omega} \left[
\frac{2 a (\cos \kappa-1)}{v_g^2-a}+ \cos \kappa-1 + \frac{\sin^2 \kappa}{(a-1)(\cos \kappa-1)} \right], \nonumber
\eea
where $v_g = d \omega / d \kappa$ is the group velocity corresponding to the dispersion relation (\ref{jjjj}).

\subsection{Non-integrable difference-difference KdV equations}

We now present the multiscale analysis of two different non-integrable lattice KdV equations obtained by 
discretizing the continuous KdV equation. They read:
\bea
u_{n,m+1}-u_{n,m-1}&=&\frac{a}{4}\left(u_{n+3,m}-3u_{n+1,m}+3u_{n-1,m}-u_{n-3,m}\right)-\label{k1} \\
&&- \frac{b}{2}\left(u_{n+1,m}^2-u_{n-1,m}^2\right), \nonumber 
\eea
and 
\bea
u_{n,m+1}-u_{n,m-1}&=&\frac{a}{4}\left(u_{n+3,m}-3u_{n+1,m}+3u_{n-1,m}-u_{n-3,m}\right)-\label{k2} \\
&&- \frac{b}{2}\left(u_{n+1,m}^2-u_{n,m}^2\right), \nonumber 
\eea
where $a,b$ are  real parameters. Clearly, the difference between Eqs.~(\ref{k1}) and (\ref{k2}) lies just in their nonlinear
part. Precisely, Eq.~(\ref{k1}) has a symmetric nonlinear part, while it is asymmetric in Eq.~(\ref{k2}).
Since the linear part of both Eqs.~(\ref{k1}) and (\ref{k2}) is the same, they have the same dispersion relation, given by
\beq \nonumber 
\omega(\kappa) = \arcsin (a\sin^3 \kappa).
\eeq

Let us start with the analysis of Eq.~(\ref{k1}) by looking at  small amplitudes solutions $u_{n,m}=\epsilon w_{n,m}$ and considering the expansion (\ref{du}). 

At $\CO(1)$ we get a set of determining equations which give
$w_0^{(\alpha)}=0$ for $|\alpha|>1$, while the $\CO(\epsilon)$ gives $w_0^{(0)}=0$ and the following determing equations:
\bea
&&( v_g \delta_{n_1}+  \delta_{m_1} )w_0^{(1 )}=0, \label{equ11a}\\
&& w_1^{(2)} = \sigma_1 (w_0^{(1)})^2,\label{equ22b}
\eea
where
\beq
v_g= \frac{d \omega}{d \kappa},
  \qquad \sigma_1=- \frac{b \cos \kappa }{2 a \sin^2 \kappa (4 \cos^3 \kappa - \cos \omega)} \nonumber. \nonumber
\eeq
As in the previous examples, Eq.~(\ref{equ11a}) is solved by $w_0^{(1)}(n_1,\{m_i\}_{i=1}^K)=w_0^{(1)}(n_2,\{m_i\}_{i=2}^K)$ with
$n_2 = n_1- v_g m_1$. Moreover, 
higher harmonics imply that $w_1^{(\alpha)}=0$ if~$|\alpha| \geq 3$. 

At $\CO(\epsilon^2)$ we get
\beq
w_1^{(0)}= \sigma_2 |w_0^{(1)}|^2, \qquad \sigma_2 = \frac{b}{ v_g}, \label{pd2}
\eeq
for $\alpha=0$. For $\alpha=1$, after removing secularities and using Eqs.~(\ref{equ22b},\ref{pd2}) we find  the following NLS equation:
\beq
\ri \delta _{m_2} w_0^{(1)} = \sigma_3 \delta_{n_2}^2 w_0^{(1)}+ \sigma_4  w_0^{(1)} |w_0^{(1)}|^2, \label{nlsr22}
\eeq
where 
\bea
&&\sigma_3 = \frac{a \sin \kappa [3 (1 - 3 \cos^2 \kappa)- v_g^2 (1 - \cos^2 \kappa)]}{2 \cos \omega},  \nonumber \\
&& \sigma_4 = \frac{b^2}{a \cos \omega \sin \kappa} 
\left[ \frac{\cos\omega}{3 \cos \kappa} + \frac{\cos \kappa}{2 ( \cos\omega - 4 \cos^3 \kappa )} \right]. \nonumber
\eea

It is remarkable  to notice that Eq.~(\ref{nlsr22}) is an integrable NLS equation (if $\ell=\infty$) even if the starting difference-difference KdV
equation (\ref{k1}) is non-integrable. At this level  the integrability of Eq.~(\ref{nlsr22}) is definitely
due to the symmetric version of the nonlinear part of Eq.~(\ref{k1}). As a matter of fact, the same multiscale expansion
carried out on Eq.~(\ref{k2}) yields a NLS equation of the form (\ref{nlsr22}), but with complex coefficients $\sigma_3, \sigma_4$,
thus breaking its integrability property. In this latter case the coefficients $\sigma_i$, $1 \leq i \leq 4$, read
$$
\sigma_1 =\frac{b \exp (\ri \kappa) }{4 a \sin^2 \kappa (4 \cos^3 \kappa - \cos\omega)},  \qquad \sigma_2= \frac{b}{2 v_g} , \nonumber 
$$
and
$$
\sigma_3= \frac{a \sin \kappa [3 -9 \cos^2 \kappa-  v^2_g \sin^2 \kappa] }{\cos \omega},  \qquad
\sigma_4= \frac{ \ri b (\sigma_1 + \sigma_2)[1- \exp (\ri \kappa)]}{2 \cos \omega}. \nonumber
$$

\section{Multiscale expansion of linearizable discrete equations} \label{sec2}

In this Section we  shall consider difference-difference and differential-difference equations, which are linearized by a transformation of coordinates,
usually a contact transformation. In particular, 
we analyze  a differential-difference Burgers
equation and  the Hietarinta equation.
Let us here recall that a multiscale reduction of the Hietarinta equation has been already considered
in \cite{lp} for $\ell=2$, while its linearizability has been proven in
\cite{nalini}.

\subsection{Discrete Burgers equations} 

We  look for a discrete Burgers equation which has a real dispersion relation. This request is equivalent to say
that its linear part admits dispersive wave solutions.
A discrete Burgers equation has been derived in \cite{hlw}, but it is not dispersive since it involves an
asymmetric discrete-time derivative. One can prove that a dispersive completely discrete Burgers-type equation which is
linearizable via a discrete Cole-Hopf map does not exist. Roughly speaking this is related to the fact that integrability implies 
non-symmetric discrete-time derivatives, which are usually incompatible with real dispersion relations.

Therefore we shall shall consider the following
dispersive differential-difference Burgers equation \cite{lrb}:
 \bea \label{b1}
 \ri a^2 \dot u_n = (1 +a u_n) ( u_{n+1}-u_n) + \frac{u_{n-1}-u_n}{1 + a u_{n-1}},
 \eea
with  $a \in \mathbb{R}$. Eq. (\ref{b1}) is obtained as a compatibility condition for the
following Lax pair:
 \beq \label{blax}
 \Psi_{n+1}(t) = (1 + a u_n) \Psi_n(t),
 \qquad
\ri  \dot \Psi_n(t) =\frac{u_n-u_{n-1}+au_n u_{n-1}}{a( 1 + a u_{n-1})} \Psi_n(t).
 \eeq
 
As Eq. (\ref{b1}) has a real dispersion relation we can  use the same procedure described in Section \ref{intro}, but with a continuous-time variable.
The linear part of Eq. (\ref{b1}) admits wave solutions with a
 dispersion relation given by
\beq \nonumber
\omega(\kappa) =\frac{2}{a^2} (\cos \kappa-1) .
\eeq

For  solutions of Eq.  (\ref{b1}) with small complex amplitude  $u_{n}(t)=\epsilon w_{n}(t)$, where $w_n(t)$ is expanded according to
a natural modification of Eq. (\ref{du}), we get at $\CO(1)$ the conditions $w_0^{(\alpha)}=0$ for $\alpha>1$ and $\alpha<0$.

At $\CO(\epsilon)$ we have $w_1^{(\alpha)}=0$
for $\alpha>2$ and $\alpha<0$ and $w_0^{(0)}=0$. Moreover we get
\beq
( v_g \delta_{n_1}+  \partial_{t_1} )w_0^{(1 )}=0, \nonumber
\eeq
whose solution is $w_0^{(1)}(n_1,\{t_i\}_{i=1}^K)=w_0^{(1)}(n_2,\{t_i\}_{i=2}^K)$ with
$n_2 = n_1- v_g t_1$, $v_g = d \omega / d \kappa$, and
\beq
 w_1^{(2)} =  \frac{ a \cos \kappa  (1-e^{-\ri \kappa})}{2
\cos \kappa  - \cos(2\kappa)-1}(w_0^{(1)})^2.\label{equ22b2}
\eeq

At $\CO(\epsilon^2)$ we get $w_1^{(0)}=0$ for $\alpha=0$, while for $\alpha=1$, using Eq.~(\ref{equ22b2}), we obtain a
linear equation
\beq
\partial _{t_2} w_0^{(1)} =  \frac{\cos \kappa}{a^2} \delta_{n_2}^2 w_0^{(1)}, \label{nlsr23}
\eeq
a remainder of the fact that Eq.~(\ref{b1}) is linearizable.

Let us now  consider  the multiscale analysis of a difference-difference equation obtained through a na\"ive, but symmetric, discretization
of Eq.~(\ref{b1}). Obviously the resulting equation is not  linearizable  and it is not associated with a Lax pair of the type
given in Eq.~(\ref{blax}). It reads
\beq \label{b4}
 \frac{\ri a^2}{2b} (u_{n,m+1}-u_{n,m-1})= 
 (1 +a u_{n,m}) ( u_{n+1,m}-u_{n,m}) + \frac{u_{n-1,m}-u_{n,m}}{1 + a u_{n-1,m}},
 \eeq
 where $b\in \mathbb{R}$ is a parameter related to the time  discretization.
 The linear part  of Eq.~(\ref{b4}) admits dispersive wave solutions with
dispersion relation
 $$
 \omega(\kappa)= \frac1b \arcsin \left[\frac{2 b}{a^2} (\cos \kappa -1)\right].
 $$
 
 By applying the usual procedure, we get, at $\CO(\epsilon^2)$, the analogue of Eq.~(\ref{nlsr23}). It reads
$$
\ri  \delta_{m_2} w_0^{(1)} = - \frac{a^2 v_g [ v_g^2
(\cos\kappa-1) -\cos\kappa]}{ 2\sin \kappa}\delta_{n_2}^2 w_0^{(1)},
$$
which is again a linear equation.

\subsection{The Hietarinta equation} 

The $\mathbb{Z}^2$-lattice equation
\begin{equation} \label{h1}
\frac{u_{n,m}+e_{2}}
{u_{n,m}+e_{1}} \frac{u_{n+1,m+1}+o_{2}}
{u_{n+1,m+1}+o_{1}}=\frac{u_{n+1,m}+e_{2}}
{u_{n+1,m}+o_{1}}\frac{u_{n,m+1}+o_{2}}
{u_{n,m+1}+e_{1}}, 
\end{equation}
has been introduced by Hietarinta in~\cite{hieta}. Here, $e_i$ and $o_i$, $i=1,2$, are
real and distinct parameters.
As proven in \cite{nalini}, Eq.~(\ref{h1}) is linearizable. For our purposes
it is convenient
to reparametrize the coefficients $e_i$ and $o_i$ by means of the following transformations:
$e_i \mapsto e_i^{-1}$ and $o_i \mapsto o_i^{-1}$, $i=1,2$.

As noticed in \cite{lp} the linear part of Eq.~(\ref{h1})
has travelling wave solutions~$u_{n,m}=\exp\{\ri[\kappa n-\omega(\kappa)m]\}$ with a complex dispersion relation. Nevertheless, the constraint $o_{1}+e_{1}=o_{2}+e_{2}$ provides the real dispersion relation
\begin{equation} \nonumber
   \omega (\kappa)=2\arctan \left[\frac{a-b} {a+b}\tan \left(\frac{\kappa} {2} \right)\right].
\end{equation}
where~$a =o_2-o_1$,  $b= e_1-e_2$.

For  small amplitude  solutions $u_{n,m}=\epsilon w_{n,m}$, with $w_{n,m}$ given by Eq.~(\ref{du}) 
we obtain from Eq.~(\ref{h1}) a set of equations which, at $\CO(1)$, give the conditions
$w_0^{(\alpha)}=0$ for $\alpha>1$ and $\alpha<0$. 

 At $\CO(\epsilon)$ we find $w_{1}^{(0)}=0$ and
 \bea
&&(v_g  \delta_{n_1}+ \partial_{m_1} )w_1^{(1 )}=0, \label{equ11a2hiet}\\
&& w_2^{(2)} = \zeta_1 (w_1^{(1)})^2,\label{equ22b2hiet}
\eea
where
\beq
v_g= \frac{d \omega}{d \kappa},  \qquad 
\zeta_1= e_{1}+ b \frac{\exp( \ri \kappa) +\exp[ \ri (\kappa + \omega)] }{\exp( \ri \kappa) - \exp( \ri \omega)}. \nonumber
\eeq
Eq.~(\ref{equ11a2hiet}) is solved by $w_1^{(1)}(n_1,\{t_i\}_{i=1}^K)=w_1^{(1)}(n_2,\{t_i\}_{i=2}^K)$ with
$n_2 = n_1- v_g t_1$.

At order $\CO(\epsilon^{2})$, by taking into account Eq.~(\ref{equ22b2hiet}), we get for $\alpha=0$, 
$$
w_{2}^{(0)}=(2e_{1}+b-a)|w_{1}^{(1)}|^2,\label{Reghini}
$$
which leads, for $\alpha=1$, to the following linear equation
\beq
\ri \delta_{m_{2}}w_{1}^{(1)}=\zeta_{3}\,\delta_{n_{2}}^{\,2}w_{1}^{(1)},\qquad
\zeta_{3}= \frac12 ( \cos \kappa - \cos \omega) \sin \omega. \label{hietNN}
\eeq

In our previous work \cite{lp}, the multiscale analysis of the Hietarinta equation provided,
in the $\ell=2$ case, the following difference-difference equation:
\bea 
&&  \ri (\phi_{n_2,m_2+1} - \phi_{n_2,m_2}) +  
   c_1  ( \phi_{n_2+2,m_2} - 2  \phi_{n_2,m_2} + \phi_{n_2-2,m_2}  )  + \nonumber  \\ 
&& \qquad +\,c_2 ( \phi_{n_2+1,m_2}  - 2  \phi_{n_2,m_2} +
\phi_{n_2-1,m_2} ) +
c_3   \phi_{n_2,m_2} |{\phi}_{n_2,m_2}|^2+ \label{yy}  \\ 
&&\qquad +\, c_4   \psi_{n_2,m_2} \, {\phi}_{n_2,m_2}+
c_5   (\phi_{n_2,m_2})^2 \, \bar{{\phi}}_{n_2,m_2} = 0, \nonumber 
\eea
where $\phi=w_{1}^{(1)}$ and the function
$ \psi$ is defined through the equation
\beq
\psi_{n_2+1,m_2} -  \psi_{n_2-1,m_2}= 
c_6  [{\bar \phi}_{n_2,m_2} (\phi_{n_2+1,m_2} - \phi_{n_2-1,m_2}) + 
\phi_{n_2,m_2} ( {\bar  \phi}_{n_2+1,m_2} - {\bar \phi}_{n_2-1,m_2} ) ]. \label{nnn}
\eeq

The coefficients $c_i$, $1 \leq i \leq 6$,   have been determined and they are all real.
Eq.~(\ref{yy}) is a (non-integrable) discrete integral equation due to the presence of  the function $ \psi $.
 It is remarkable to note that, if $\ell=\infty$, Eq.~(\ref{nnn}) can be explicitly integrated and the 
resulting coefficients $c_i$  combine in such a way that the nonlinear term in Eq.~(\ref{yy}) vanishes, thus giving Eq.~(\ref{hietNN}).

\section{Concluding remarks} \label{sec4}

In this paper we presented a Conjecture which gives necessary conditions for
the integrability and linearizability of dispersive difference-difference and
differential-difference equations. This Conjecture has been confirmed by a
long list of examples, contained in Sections \ref{sec1}, \ref{sec2}. In particular,
performing the discrete multiscale analysis, we found that the resulting NLS equation turns out to be integrable if:

\begin{enumerate}

\item the starting  $\mathbb{Z}^2$-lattice equation is integrable, or
\item the starting  $\mathbb{Z}^2$-lattice equation is non-integrable, but obtained through a symmetric discretization of a continuous integrable
equation.

\end{enumerate}

We expect that claim (2) fails at higher orders of the multiscale expansion. 
Moreover 
the resulting NLS equation turns out to be non-integrable if
 the starting nonlinear $\mathbb{Z}^2$-lattice equation is obtained through a non-symmetric discretization of a continuous integrable
equation. Let us recall that  discrete symmetric derivatives play a fundamental role in the integrability of discrete equations, see
 \cite{yamilov}. We stress here that the integrability of the NLS equation is
 lost as soon as we go to a finite slow-varyness order since shifts operators
 do not satisfy the Leibniz rule and thus the reduced Lax pair is not
 compatible anymore.

Concerning the linearizability of $\mathbb{Z}^2$-lattice equations we obtained a linear Schr\"odinger equation if:

\begin{enumerate}

\item the starting $\mathbb{Z}^2$-lattice equation is linearizable, or
\item the starting  $\mathbb{Z}^2$-lattice equation is obtained through a discretization of a continuous linearizable
equation.

\end{enumerate}

The above characterization is evident just by considering 
expansions on functions of infinite order of slow-varyness. As
soon as we choose a finite order, as shown for
$\ell=2$, we may obtain nonlocal discrete equations as the
integration of the equations defining the auxiliary harmonics may
not be exact \cite{lp}.

This work still leaves many open problems. Work is in progress on:
\begin{itemize}
\item The proof of the Conjecture for some relevant classes of lattice
  equations. 
\vspace{.2truecm}
\item The construction of higher orders in the multiscale expansion and
 on the definition of an order of integrability of a
  lattice equation (in the spirit of the paper \cite{dp}).
\end{itemize}
It is worthwhile to notice that a meaningful classification of integrable $\mathbb{Z}^2$-lattice equations can be obtained only
   by going at least to the fourth order in the multiscale expansion (indeed, the third order is not enough 
   to show the non-integrability of the discretized KdV equation (\ref{k1}))

In the present paper we have considered the modulation of weak plane waves solutions 
of the discrete systems.  Work is also in progress on the modulation of 
constant solutions. In such a case the multiscale analysis  should provide
a KdV-type equation instead of a NLS one \cite{LevHer}.

\section*{Acknowledgments}
DL,  MP and CS were partially supported by PRIN Project {\it Metodi geometrici nella teoria delle onde non lineari ed applicazioni-2006} of the Italian Minister
for Education and Scientific Research.
RHH was partially supported by the Region of Madrid and Universidad Polit\'ecnica de Madrid (UPM) with grant CCG--UPM/MTM--539 and the Spanish Ministry of Science, Project MTM 2006--13000--C03--02.

\end{document}